\documentclass[aps, prl, twocolumn,showpacs,preprintnumbers,amsmath,amssymb,nofootinbib]{revtex4-1}

\usepackage{graphicx}
\usepackage{dcolumn}
\usepackage{bm}

\newcommand\ignore[1]{}
\def\one{{\,\hbox{1\kern-.8mm l}}}

\def\Tr{{\rm Tr\, }}

\newcommand{\Cset}{{\,\,{{{^{_{\pmb{\mid}}}}\kern-.45em{\mathrm C}}}}}

\newcommand{\be}{\begin{equation}}
\newcommand{\ee}{\end{equation}}
\newcommand{\bea}{\begin{eqnarray}}
\newcommand{\eea}{\end{eqnarray}}

\def\b{\beta}

\begin{document}

\title{Holography for the very early Universe  and the classic puzzles of Hot Big Bang cosmology}

\author{Horatiu Nastase$^{1}$}\email{horatiu.nastase@unesp.br}
\author{Kostas Skenderis $^{2}$}\email{K.Skenderis@soton.ac.uk}
\affiliation{${}^{1}$Instituto de F\'{i}sica Te\'{o}rica, UNESP-Universidade Estadual Paulista, Rua Dr. Bento T. Ferraz 271, Bl. II, 
S\~ao Paulo 01140-070, SP, Brazil}
\affiliation{$^{2}$STAG Research Center and Mathematical Sciences, University of Southampton, Highfield, Southampton SO17 1BJ, United
Kingdom}
\date{\today}

\begin{abstract}
We show that  standard puzzles of hot Big Bang cosmology that motivated the introduction of cosmological inflation, such as the smoothness and horizon problem, the flatness 
problem and the relic problem % the way to generate enough entropy and perturbations, %and baryon asymmetry, 
are also solved by holographic models for very early universe 
%of \cite{McFadden:2009fg}, which are 
based on perturbative three dimensional QFT. % representing a non-geometric early Universe. 
In the holographic setup, cosmic evolution is mapped to inverse renormalization group (RG) flow of the dual QFT, and the resolution of the puzzles relies on properties of the RG flow. 
%These considerations, together with the fact that these models provide as good a fit to CMBR data as inflation,  put them on equal footing  with  inflation, as 

\end{abstract}

%\pacs{}

\maketitle

%==============================================================================
%\section{Introduction}
%==============================================================================

The theory of inflation was initially introduced \cite{Guth:1980zm, Linde:1981mu, Albrecht:1982wi}   as an answer to 
three problems of hot Big Bang cosmology : (i) the horizon problem (why is the universe so homogenous despite the fact that separated regions were causally disconnected?), (ii) the flatness problem (why is the Universe as flat as we see it today),  and (iii) the relic problem (why we do not see any relics from the very early Universe?).  Inflation beautifully resolves  these (and other) problems by postulating a period of accelerating expansion in the very early Universe.

What is perhaps the biggest success of this theory is its ability to generate primordial perturbations, which form the seeds for structure formation in the late Universe, and which are in excellent agreement with observations of the cosmic microwave background (CMB)  by satellites and other missions. Despite these successes, however, the underlying theory still remains unsatisfactory: it requires fine tuning, there are trans-Planckian issues and questions about initial condition, see for example \cite{Brandenberger:1999sw}. The theory of inflation is an effective field theory and we are still lacking a proper understanding of its ultraviolet (UV) completion. This as well as the resolution of the initial singularity require the embedding of inflation in a consistent quantum theory of gravity. Achieving  such embedding in string theory is an on-going effort and the very existence of (quasi)-de Sitter solutions in string theory has recently been questioned (see for example \cite{Obied:2018sgi}). It is thus important to approach this question from different perspectives and explore and further develop alternative models for the very early Universe. 

It is widely believed that quantum gravity is holographic 
\cite{tHooft:1993dmi, Susskind:1994vu, Maldacena:1997re}, meaning that there is an equivalent description using a quantum field theory (QFT) (with no gravity) in one dimension less.
Holographic dualities are still conjectural and this is even more so in the case of cosmology.  The cosmological holographic framework however has already passed a number of non-trivial tests and we will provide additional support in this paper. Work on holographic cosmology was initiated in \cite{Witten:2001kn, Strominger:2001pn, Strominger:2001gp, Maldacena:2002vr}, with 
standard inflation fitting in this framework as a strongly coupled QFT (see for example \cite{Maldacena:2011nz, Hartle:2012qb, Hartle:2012tv,Schalm:2012pi, Bzowski:2012ih, Mata:2012bx, Garriga:2013rpa, McFadden:2013ria, Ghosh:2014kba, Garriga:2014ema, Kundu:2014gxa, Garriga:2014fda, McFadden:2014nta, Arkani-Hamed:2015bza, Kundu:2015xta, Hertog:2015nia,Garriga:2015tea,  Garriga:2016poh,Hawking:2017wrd,Arkani-Hamed:2018kmz}). Holographic cosmology  also contains 
qualitative new models for the very Early Universe obtained by 
considering QFTs at weak coupling  \cite{McFadden:2009fg}. These new models correspond to a strongly-coupled non-geometric phase of gravity and they will be the focus of this paper. 

In the context of cosmology  the dual QFT is a three dimensional Euclidean theory, which is located at future infinity and its partition function, in the presence of sources for gauge invariant operators,  is identified with the wavefunction of the universe. The fields parametrizing the (Dirichlet)  boundary conditions at future infinity\footnote{Note that the asymptotic structure near the  timelike boundary of AdS \cite{FG,deHaro:2000vlm} is mapped to the asymptotic structure near the spacelike boundary of de Sitter \cite{Starobinsky:1982mr} via analytic continuation \cite{Skenderis:2002wp}, see also \cite{Mazur:2001aa} and \cite{Poole:2018koa}. The same analytic continuation also maps general perturbations (at least to quadratic order) around Domain-Walls/FRW cosmologies \cite{McFadden:2009fg,McFadden:2010na, McFadden:2010vh,McFadden:2011kk}, and this translates into specific analytic continuation on the QFT side, as discussed in \cite{McFadden:2009fg,McFadden:2010na, McFadden:2010vh,McFadden:2011kk}.}  are identified with the sources of dual operators and the arguments of the wavefunction of the Universe \cite{Maldacena:2002vr}. 
The dimension which is reconstructed holographically is the time direction and 
cosmic evolution is mapped to inverse RG flow. The holographic description is currently known only for the very Early Universe, the period usually associated with inflation, and the transition to standard cosmology  is via ``instant reheating'', {\it i.e.} the outcome of this period becomes the initial conditions for the subsequent evolution via Einstein equations (see \cite{Easther:2011wh}).

In the holographic framework, models are defined by providing the dual QFT, and in the models
describing a non-geometric early Universe this is a three dimensional super-renormalizable theory: $SU(N)$ gauge theory for a gauge field $A_i$ coupled to
 scalars $\phi$ and fermions $\psi$, with action \cite{McFadden:2009fg}
\bea \label{QFTaction}
S&=&\frac{1}{g^2_{YM}}\int d^3x {\rm Tr}\left[\frac{1}{2}F_{ij}F^{ij}+(D\phi)^2+\bar \psi D_i \gamma^i \psi
\right.\cr
&&\left.
+ \mu (\bar \psi \psi \phi)+\lambda \phi^4\right]\;,
\eea
plus a non-minimal coupling $\int d^3x \xi R \phi^2$, where $R$ is the scalar curvature -- on a flat 3d background the non-minimality parameter $\xi$ appears only in the improvement term in the energy momentum tensor. All fields are in the adjoint of $SU(N)$ and we suppress numerical factors and flavour indices (see \cite{Afshordi:2016dvb} for the details).
This theory has  a %the property of 
``generalized conformal 
structure'', which means that if one promotes $g^2_{YM}$ to a field with appropriate conformal transformations 
the theory becomes conformal  \cite{Jevicki:1998yr, Kanitscheider:2008kd}, or that by assigning 
``4d dimensions'' to the fields, $[\Phi]=[A_i]=1$, 
$[\psi]=3/2$, all terms in the action scale in the same way. 

The phenomenology of these models has been worked out in \cite{McFadden:2009fg, McFadden:2010na, McFadden:2010vh, McFadden:2011kk, Bzowski:2011ab, Coriano:2012hd, Kawai:2014vxa,McFadden:2010jw}, using methods from 
\cite{Skenderis:2002wp,Papadimitriou:2004ap, Papadimitriou:2004rz, Maldacena:2002vr}.
The models predict a scalar power spectrum of the form 
\be \label{power_spec}
\Delta_S^2=\frac{\Delta_0^2}{1+\frac{g q^*}{q}\ln \left|\frac{q}{\b g q^*}\right|
+{\cal O}\left(\frac{gq^*}{q}\right)^2}\;,
\ee
where $\beta, g$ are parameters that are obtained by a 2-loop computation of the 2-point function of the energy momentum tensor and
there is a similar form for the tensor power spectrum. 
%The model also predicts non-Gaussianity which is of the exactly factorisable equilateral shape with $f_{NL}^{\rm equil}=5/36$ \cite{McFadden:2010vh}.
These models have been compared against WMAP \cite{Dias:2011in, Easther:2011wh} and
 Planck data \cite{Afshordi:2016dvb,Afshordi:2017ihr}
and it was found that within their regime of validity\footnote{One of the results of \cite{Afshordi:2016dvb,Afshordi:2017ihr} is 
that the model becomes non-perturbative at very low multipoles (less that 30) and a non-perturbative evaluation of the power 
spectrum is needed to model this region.} they provide an excellent fit to data and are competitive with 
$\Lambda$CDM  -- the fit to data shows that holographic cosmology (HC) and $\Lambda$CDM are within one sigma.

In this Letter, we would like to discuss how holographic cosmology addresses the hot Big Bang problems. We will start by first reviewing how inflation solves these problem and then 
move to discuss them within the context of holographic cosmology. As this part is standard material we will be brief -- the details can be found in most cosmology textbooks.
\paragraph{\bf Inflation and hot Big Bang problems}--

1. {\em Smoothness and horizon problems}, or {\em Why is the Universe uniform and isotropic?}

The question can be formulated as follows: why is the Universe so smooth and correlated on large scales %, as shown for instance in the CMBR data, 
when different parts of the sky %, for instance CMBR light coming from opposite parts of the sky, 
were not in causal contact at the initial time? In hot Big Bang cosmology, the points in the CMB separated by more than $1.6^\circ$ could not have been in causal contact  because their past light cones do not overlap before the spacetime is terminated by the initial singularity (see for example \cite{Weinberg:2008zzc, Baumann:2014nda}). One has to increase the horizon distance at the surface of last scattering at least by a factor of 66 to be consistent with observations. 

%A simple estimate (see for example \cite{Weinberg:2008zzc}) shows that in Hot Big Bang cosmology no physical influence could have smoothed   initial inhomogeneities.

%If $d_H(t_0)$ is the horizon distance at the time of last scattering (CMBR creation) 
%and $r_H(t_0)$ is the distance travelled by light from last scattering until today in today's scales, then a short computation shows that 
%$
%N=2r_H(t_0)/d_H(t_0) \simeq  72$
%%2\left(\frac{t_0}{t_{\rm ls}}\right)^{1/3}=2\left(\frac{a_0}{a_{\rm ls}}\right)^{1/2}\cr
%%2(1+z_{\rm ls})^{1/2}\simeq 72$, 
%and $d_H(t_0)$ should be increased by at least that much 
%%one has to increase by at least as much the horizon distance at the surface of last scattering in order 
%to be consistent with observations. 

Inflation's answer  to this problem is that the exponential blow up of a small patch creates the whole observable Universe, 
and %at the beginning 
this patch was in causal contact. Let $t_{\rm bi}$ the time inflation began, $t_I$ it ended and  $N_e=H(t_I) (t_I-t_{\rm bi})$ the number of e-foldings. Assuming nothing much happened between the end of inflation and the beginning of radiation domination, a short computation shows that the horizon problem is avoided provided we have enough e-foldings of inflation.
%%\begin{equation} \label{condition}
%$e^{N_e}\gsim e^{56}\frac{\rho_1^{1/4}}{5\times 10^{13}GeV}$,
%%\end{equation}
%where $\rho_1$ is the energy density at the beginning of the radiation-dominated era.

%Quantitatively, we have exponential inflation from $t_{\rm bi}$ to $t_I$,
%with the number of e-folds of inflation being $N_e=H_I (t_I-t_{\rm bi})$, and find
%\be
%d_H(t_{\rm ls})\simeq \frac{a(t_{\rm ls})}{a(t_I)}\int_{t_{\rm bi}}^{t_I}dt\; e^{H_I(t_I-t_{\rm bi})}\simeq \frac{a(t_{\rm ls})}{a(t_I)H_I}e^{N_e}\;,\label{dh}
%\ee
%where $t_I$ is the time when inflation ends,
%whereas the distance travelled by light from last scattering to now, but measured at last scattering, is $
%r_H(t_{\rm ls})=a(t_{\rm ls})\int_{t_{\rm ls}}^{t_0}\frac{dt'}{a(t')}\simeq \frac{2a(t_{\rm ls})}{a_0 H_0}$, 
%leading to the condition %that $d_H(t_{\rm ls})/2r_H(t_{\rm ls})$ as
%$e^{N_e}> \frac{a(t_I)H_I}{a_0 H_0}$,
%which is %found to be 
%\be
%e^{N_e}\gsim e^{56}\frac{\rho_{\rm begRD}^{1/4}}{5\times 10^{13}GeV}.\label{condition}
%\ee

2. {\bf Flatness problem}, or {\em Why do we have $\Omega\simeq 1$ in the past?}

Observations tell us that the Universe is approximately flat today. If the Universe were exactly flat in the past, then cosmic evolution would preserve this property and it would be exactly flat today.
However, if $\Omega -1 \neq 0$ but small, extrapolating into the past using matter domination (MD)
and radiation domination (RD)  formulae, we find an extremely flat Universe at initial times. Quantitatively, 
$\Omega(t)-1\propto t^{2(1-p)}$
%\be
%\Omega(t)-1=\frac{k}{a(t)^2H(t)^2}\propto \left(\frac{t}{a(t)}\right)^2\propto t^{2(1-p)}\;,
%\ee
for $a(t)\propto t^p$. In both RD ($p=1/2$) and MD ($p=2/3$) eras, $\Omega(t)-1$ increases with time, so it must have been very small in the past, and to avoid fine tuning we need a period of $p>1$, 
to bring it down to the value we obtain now. 

Indeed, inflation naturally drives $\Omega$ very close to one. A short computation (see for example  \cite{Weinberg:2008zzc})
shows that $ \Omega_0-1
%&=&\frac{k}{a_0^2H_0^2}=\frac{k}{a_{\rm bi}^2H_{bi}^2}e^{-2N_e}\left(\frac{a(t_I)H_I}{a_0 H_0}\right)^2\cr
=(\Omega(t_{bi})-1)
e^{-2N_e}\left((a(t_I)H(t_I)/(a_0 H_0)\right)^2$,
where (as usual) the subscript 0 denotes todays values, and this leads to exactly the same condition 
needed to solve the horizon problem.
%(\ref{condition}).
%To put some numbers in, using RD evolution, we find that $\Omega-1\sim 10^{-16}$ at $e^+e^-$ annihilation, and then at the (end of the)
%inflationary time we have $(\Omega-1)_I\sim 10^{-54}$,
%%\be
%%(\Omega-1)_I=(\Omega-1)_{e^+e^-}\left(\frac{a_e H_e}{a_I H_I}\right)^2
%%%=10^{-16}\left(\frac{T_e}{T_I}\right)^2
%%\sim 10^{-54}\;,\label{fluct}
%%\ee
%where we have used $T_{e^+e^-}\sim 1 MeV$, $T_I=T_{\rm inflation}\sim 10^{16}GeV$.

3. {\bf Relic (monopole) problem}, or {\em Why don't we see relics in the Universe?}

In phase transitions we would obtain relics, for example monopoles from GUT phase transitions, 
where we would expect about one monopole 
per nucleon, or $10^{-9}$ monopoles per photon. However, from direct searches in materials on Earth 
 we know that there are $\leq 10^{-30}$ monopoles per nucleon (see \cite{Weinberg:2008zzc}, chapter 4.1.C), %or $10^{-39}$ monopoles per photon, 
so we need a reduction factor of $10^{-30}$. 

Relics in general are also constrained by their gravitational effects
(see \cite{Kolb:1990vq}, chapter 7.5): in order to not over-close the Universe, we need a reduction factor of 
about $10^{-11}$, much less stringent than for monopoles.

Inflation's answer to the problem of both monopoles and general relics 
is that it dilutes them % monopoles, created as one per horizon, 
during the period of exponential inflation. Inflation therefore needs to happen 
after, or at most during the phase transition. %Quantitatively, we need

\paragraph{\bf Resolution using holographic cosmology}---

We now turn to the same questions in the context of holographic cosmology, where gravity is strongly coupled and the dual field theory is weakly 
coupled. 

1. {\bf Smoothness and horizon problems}

%As reviewed, the horizon problem originates from the fact that various region that 
%appear casually connected, they could not have been so in hot Big Bang cosmology because the spacetime terminates by the initial singularity. %In inflation, the problem is resolved using an exponential expansion before hot Big Bang.

These models describe a non-geometric early Universe so geometric concepts such as light-cones are meaningless, and the traditional formulation of the problem is not valid. Nevertheless it would be useful to understand the mechanism that put in causal contact parts of the sky that from the perspective of Hot Big Bang cosmology appear to be uncorrelated.

In holographic cosmology cosmological observables are computed from correlation functions of the dual QFT, and the correlations at the surface of last scattering are those of these correlators. 
In QFT correlation functions at different scales are related to each other via renormalization group flow. As time evolution is mapped to inverse RG flow, points widely separated at the surface of last scattering would be linked by RG flows that connect the UV with the deep IR, so as long as the QFT is well-defined in the IR, there is no horizon problem, as any two points at the surface of scattering will be causally linked via a deep enough RG flow.

%The deep IR of the QFT is linked to the earliest times. 
The theories we discuss here are super-renormalisable so they are naively IR divergent. This is the holographic dual of the bulk initial singularity. These class of theories however are expected to be non-perturbatively IR finite \cite{Jackiw:1980kv, Appelquist:1981vg} and this has recently been confirmed  by lattice studies
\cite{Lee:2019zml}. It follows that in this class of models there is no horizon problem, irrespectively of the details of each model.

We now illustrate that the usual inflationary resolution of the horizon problem
is an example of the same mechanism. For concreteness we discuss the case of asymptotically de Sitter inflation but the same comments apply more generally. % and take $t=0$ to be the end of this phase. 
Any two points separated by distance $L$ at the space-like boundary  of de Sitter (the end of this phase) may be linked via bulk  geodesics that go to the interior of de Sitter. From the perspective of the dual QFT (and after using the domain-wall/cosmology correspondence \cite{Skenderis:2006jq} to map this question to AdS), the (renormalised) length of these geodesics provide the 2-point function of a dual operator inserted at each of the two points \cite{Graham:1999pm}. A short computation (see for example \cite{Maldacena:1998im,Rey:1998ik}) shows that 
$L\propto 1/r_0$, where $r_0$ is the maximum radial distance reached in the bulk. Recall that the radial  coordinate encodes RG flow in the dual QFT, so the number e-folding corresponds to the amount of RG flow for which the dual field theory is strongly coupled and nearly conformal: %description should be valid, i.e. how early the inflationary phase started relative to the end of this period,
% If the theory is IR finite 
 %(corresponding to a number of e-folds during which we must 
%have inflation in the standard scenario), 
%after which usual radiation dominated cosmology can follow. 
it is simple to verify (using the fact that $L\propto 1/r_0$) that the factor multiplying the RG scale corresponds to the factor $e^{N_e}$ in inflation. %So if we know the QFT 

%If this constraint holds then knowledge of the QFT over these energy scales only is enough to resolve the horizon problem.

%Phrased in this way the requirement applies equally well to the geometric as well as the non-geometric phase. 

2. {\bf Flatness problem}

To formulate the question in the context of holographic cosmology, 
we consider a small deviation from a flat  background ($\Omega=1$)
%in the cosmology and associated domain wall 
and we would like to show that under time evolution (=inverse RG flow) the flat geometry is an attractor. 
Like in the inflationary case, this needs to be addressed independently of the usual cosmology that follows: 
we must show that the {\em non-geometric phase alone} can do this.

In holography the spacetime where the dual QFT lives is a fixed non-dynamical background, so one may 
wonder whether the flatness problem makes sense in this context. 
%The dynamics of the QFT will not change the spacetime the QFT is formulated on: if the QFT is on flat background will remain on a flat background. %The flatness problem arises when we consider small deviations from flatness, and this problem has a formulation in holography. 
A small deviation from flatness means that the spacetime metric is $g_{ij} = \delta_{ij} + h_{ij}$, where  $\delta_{ij}$ is 
the metric on flat $\mathbb{R}^3$ and $h_{ij}$ is a small deviation. By a standard argument, the deviation induces a new coupling in
the action $\int d^3 x T^{ij} h_{ij}$, where $T_{ij}$ is the energy-momentum tensor of the dual QFT on 
$\mathbb{R}^3$ (plus higher order terms). The new coupling $h_{ij}$  will run under RG flow and (as time evolution is inverse RG flow)
this is the counterpart of the fact that the density parameter $\Omega$ evolves in a non-flat FRLW. Note that if $h_{ij}=0$, 
this coupling will not be induced by the RG flow (in a Lorentz invariant QFT) and this is counterpart of the statement that
if the Universe is flat, it remains flat at all times.

The flatness question is now whether the new coupling dies off or dominates in the UV. If it dies off then the flat geometry is an attractor, as in inflation. The perturbative superrenormalizable QFTs (with action given in (\ref{QFTaction})) that feature in holographic cosmology have a generalized conformal structure and this implies that when the coupling is very small they  effectively behave like CFTs: they are nearly conformal. Since we are interested in the late time behaviour and the QFT is super-renormalizable, there is no loss of generality to assume that we are in the regime where the QFT is nearly conformal.
The question is then whether the deforming operator ({\it i.e.} $T^{ij}$) is relevant or irrelevant. 
Since the deformation is also assumed to be very small in the UV ($\Omega \sim 10^{-54}$), it suffices 
to compute the dimension of $T$ in the undeformed theory  \footnote{In the deformed theory, the leading correction can be computed using conformal perturbation theory and it is of order ${\cal O}(h_{ij}^2$)}.  If the operator is relevant it would die off in the UV and dominate in the IR and the opposite if it is irrelevant.

We therefore need to determine the dimension of $T$ and this can be done from its 2-point function.
In momentum space (and close to the fixed point) the 2-point function should behave as 
$q^{2 \Delta -d}$ (see for example \cite{Bzowski:2013sza}) and we can extract $\Delta$ from it.
 $T_{ij}$ is of course marginal (dimension
3 in 3 dimensions) at the classical level, and at the quantum level the $\langle T_{ij} T_{kl}\rangle $ correlator decomposes into 
a scalar and a tensor piece, both of the type $q^3N^2f(g^2_{\rm eff})$, where $g^2_{\rm eff}=g^2N/q$ is dimensionless. The factor of 
$q^3$ captures the classical dimension of $T$ and implies that to leading order the CMBR power spectra are scale invariant. In perturbation theory, $g^2_{\rm eff}\ll 1$, and at 2-loops (see \cite{McFadden:2010na, Easther:2011wh, Afshordi:2017ihr} for details) the form $f$ is\footnote{Note that as the theory (\ref{QFTaction}) is asymptotically free, the two point function $\langle T_{ij} T_{kl}\rangle $ in the underfomed theory approaches its free-field value as $q \to \infty$ (and thus $g^2_{\rm eff} \to 0$), 
{\it i.e.} all loop corrections vanish. Here we are interested to extract the precise way these corrections 
go to zero, as this controls how the new coupling behaves in the deformed theory.}
\be
f(g^2_{\rm eff})=f_0\left(1-f_1g^2_{\rm eff}\ln g^2_{\rm eff}+f_2 g^2_{\rm eff}+{\cal O}(g^2_{\rm eff})\right)\;,
\ee
where $f_1<0$ both for the best fit to the CMBR data, and for most of the general theoretical parameter space. %(see \cite{Afshordi:2016dvb} for more details), which is valid at $g^2_{\rm eff}\ll 1$, and means that 
This implies (again for $g^2_{\rm eff}\ll 1$) that  $f(g^2_{\rm eff})\propto q^{2\delta} \sim 
%1+2\delta \ln q\sim 
1-2\delta \ln g^2_{\rm eff}+...$ giving $2\delta \simeq f_1 g^2_{\rm eff}<0$,
and thus $\Delta = 3+ \delta$ making $T_{\mu\nu}$ (marginally) relevant \footnote{Note that in the standard CMBR inflationary description, $f_1 <0$  translates into a red tilt ($n_s-1<0$). Turing things around flatness implies that the spectrum should be red.}. This means that the perturbation will die off in the UV and it would lead to changes of order one in the IR. Recalling that in holographic cosmology  time evolution corresponds to inverse RG flow, i.e. from IR to UV, this is precisely what we set out to show.

%We can then turn the constraint on the flatness of CMBR fluctuation spectrum into a selection tool for field theories and their RG flows: 
%what field theories have the energy-momentum tensor as relevant operator (we saw that the class of superrenormalizable theories does for most of the theoretical parameter space), 
%and how much RG flow do we need to go from an $10^{-54}$ coefficient to one of order one (this will be a constraint on the amount of RG flow, 
%dual to the time period in cosmology)?
%This is then the equivalent in inflation on the constraint on the number of e-folds (\ref{condition}).

%In the holographic picture, time evolution corresponds to inverse RG flow, i.e. from IR to UV, and we have shown that a small 
% 
%
% what we need to show is that for this flow from IR to UV, 
%a small {\em gravitational } perturbation, deforming space from flat, is very small in the UV (for our inflationary example, $10^{-54}$ for our UV scale
%of $10^{16} GeV$, which 
%corresponds to the "end of inflation", and that we can "see" 
%at large angular momentum $l$'s in the CMBR data), 
%yet becomes of order one in the IR, corresponding to the "beginning of time". But that amounts, in  the field theory picture, to saying that 
%even a $10^{-54}$ small deformation in the UV by the energy-momentum tensor will leads to a nontrivial IR field theory. The energy-momentum tensor
%must then be a (marginally) relevant operator, that takes us away from the naive IR theory, despite its infinitesimal coefficient. 

3. {\bf Relic and monopole problem}

Let us start with monopoles. To study this problem our starting point should be a bulk theory with GUT phase transition 
and analyse how the effects of monopoles are encoded in the dual QFT.  To avoid the monopole problem we need to 
establish that their effects are washed out at late times, or equivalently in the UV from the perspective of the dual QFT.
 
Bulk gauge symmetries correspond to boundary global symmetries, so to properly analyse this problem we would need to consider boundary QFTs  that have the required global symmetry and pattern of symmetry breaking. It is an interesting problem (that we leave for future work) to classify the QFTs with such properties, start with 't Hooft-Polyakov monopoles in the bulk and analyse their effects in complete generality. 

Here we will proceed by solving a related  problem: we will consider instead a Dirac monopole in the bulk. The bulk theory will thus involve a $U(1)$ gauge field and we should consider a monopole field $A_\mu$ in the bulk, which by the standard AdS/CFT dictionary, will induce a new coupling in the boundary theory, $\int d^3 x A_{(0)i}  \tilde{j}^i$, where $\tilde{j}^i$ is the magnetic current and $A_{(0)i}$ is the boundary value of $A_\mu$.  As in our study of the flatness problem, we would like to 
study whether such a coupling will have an effect in the UV, and this can be analysed by extracting 
the dimension of $\tilde{j}^i$  near the UV when the theory is nearly conformal.

This is still a non-trivial problem as we usually work with electric variables. Luckily, 3d CFTs with a global $U(1)$ symmetry allow for an $Sl(2;\mathbb{Z})$ action, whose $S$-generator 
exchanges the electric and magnetic currents \cite{Witten:2003ya}.
%particle-vortex duality in 3d exchanging the electric current $j_i$ with the topological current $\tilde j_i=\frac{1}{2\pi}\epsilon_{ijk}\d_j A_k$, with $A_k$ the source for $j_k$  . 
In the bulk this operations corresponds to to usual electromagnetic duality (see also \cite{Herzog:2007ij,Murugan:2014sfa}). 
The 2-point function of symmetry currents in a CFT is given by (ignoring the contact terms which 
are relevant in general for the
action of $Sl(2;\mathbb{Z})$ but not relevant for us)
\be \label{2-pt}
\langle j_i (q) j_j(-q)\rangle\simeq
q \left(\delta_{\mu\nu}-\frac{q_\mu q_\nu}{q^2}\right) t
\ee
where $t$ is a constant (in a CFT) and the $S$-generator takes $t \to 1/t$. This is not a symmetry:  it maps one CFT to another. In a theory with a generalised conformal structure the form of the 2-point function is the same but $t$ is now a function of $g^2_{\rm eff}$. We will assume that the discussion in 
\cite{Witten:2003ya} generalises to such theories, at least when $g_{\rm eff} \ll 1$ and the theory is nearly free (and thus nearly conformal).

Our strategy is now to start from a theory with an electric current, compute its 2-point function to 2-loop order and then use the $S$-operation to obtain the corresponding result for the theory with the magnetic current, from which we will read off its anomalous dimension. This computation will be done in a toy model: an $SU(N)$ gauge theory 
 for a gauge field $A_i$ coupled to 
6 complex scalars $\phi^a_\alpha$, $a=1,2,3$ and $\alpha=1,2$, with the 
index $a$ transforming in the $3$ of $SO(3)$ (all fields are also in the adjoint of $SU(N)$). The Euclidean action is (we denote spatial indices by $i=1,2,3$)
\be \label{mono}
S=\frac{2}{g^2_{YM}} \int d^3x\Tr\left[\frac{1}{4}F_{ij}F^{ij}+|D_i\vec{\phi}_\alpha|^2
+\lambda|\vec{\phi}_1\times \vec{\phi}_2|^2\right]%\;,
\ee
and the global symmetry current is $
j_i^a=\sum_{\alpha=1,2}\vec{\phi}_\alpha^*T^aD_i\vec{\phi}_\alpha+h.c.$,
where $T_a$ are $SO(3)$ generators. This model has the feature of admitting Abelian vortex
solutions of the form $\phi^a_1=\phi^a_1(r) e^{i\phi}$, $\phi_2^a=\phi^a_2(r)  e^{i\phi}$, which may be used to justify the $S$-operation below, as it will be explained in detail elsewhere \cite{followup}

A 2-loop calculation, whose details will be presented in \cite{followup}, leads to the 2-point function in (\ref{2-pt}) with 
$t = 1+ 2g_{\rm eff}^2/\pi^2  \ln q$, 
%\be
%\langle j_i^a (p) j_j^b(-p)\rangle =
%\frac{p}{4}\delta^{ab}\left(\delta_{ij}-\frac{p_i p_j}{p^2}\right)\left[1+
%\frac{4}{\pi^2}\frac{g^2N}{p}\ln p\right]\;,\nonumber
%\ee
which means that the anomalous dimension of $j_i^a$ is given by $2\delta(j) =\frac{4}{\pi^2}g^2_{\rm eff}>0$, 
making $j_\mu^a$ an irrelevant operator. Applying the $S$-operation then implies $\delta(\tilde j)=-\delta (j)=-\frac{2}{\pi^2}g^2_{\rm eff}<0$. It follows that the effects of the Dirac monopole in the bulk are washed out in the UV. %Thus the theory in (\ref{mono}) does not suffer from a monopole problem. 

In general such analysis may be used to rule out holographic models: only models with negative anomalous dimension for the magnetic current solve the monopole problem.

Other relics may be studied in a similar way. We note however that the main effect is via the gravitational perturbation they generate and as such analysis will be similar to that of the flatness problem.

 {\bf Entropy problem and the arrow of time}

The current total entropy of the Universe (about $10^{88}$ per horizon volume today) requires an explanation because it is appears either too large or too low. 
Evolving to the past with standard RD and MD formulae, we find that the entropy inside the horizon at Big Bang Nucleosynthesis was $S_H(t_{\rm BBN})\sim 10^{63}$,  but one may have expected a number of order one per horizon in standard cosmology, at least at the end of a phase transition. %Inflation's answer is that we have quantum fluctuations, which then reheating transforms into a large number of photons per particle,  thus entropy per baryon. 
%The exponential expansion leads to a large volume, so a large total number of photons. In fact, the entropy per comoving volume 
%is $s\propto T^3$, for radiation $\rho_R\propto T^4$, %so $s\propto \rho_R^{3/4}$, 
%and during reheating %one finds that 
%$\rho_R\propto
%a^{-3/2}$, and so  the {\em total} entropy increases as
%$
%S\propto a^3 \rho_R^{3/4}\propto a^{15/8}$. 
On the hand, as emphasised by Penrose \cite{Penrose} (see also \cite{Carroll:2004pn, Wald:2005cb})  the entropy of the observable Universe could have been a lot higher: if the entire mass of the observable universe were collected into a single black hole the entropy would be about $10^{121}$! Usually this version is associated with the question of the arrow of time and the very special nature of the initial conditions needed in the very early Universe, and in general this issue is considered an open problem. 

In holography, time evolution is inverse RG flow, so the arrow of time is linked to the monotonicity of the RG flow, which for three dimensional QFTs was established in  \cite{Casini:2012ei}. The total entropy grows because the degrees of freedom in the UV are larger than that in the IR.
This is a general property of RG flows and not a choice of a model.  Furthermore, universality of IR dynamics makes the low entropic initial conditions natural. To explain quantitatively 
why the total entropy is as large we observe it today requires developing a holographic model for reheating and this is outside the scope of this work.

%Basically, the Standard Model particles are also part of the general field theory on the boundary, just that a part that doesn't become the gravitational sector. 

%the field theory has a very large number of degrees of freedom, which moreover 
%is {\em larger in the UV than in the IR}, which is a general statement on 
%RG flows. 
%That means that the result $S_1\sim 10^9$ is understood as being true at the end of the inflationary period in the UV, but in the 
%IR it could be close to one. Therefore {\em this is the natural expectation in field theory}, based on properties of RG flows, and not a choice of model. 

%5. {\bf Perturbations problem}
%
%This problem has already been addressed in the foundational papers on holographic cosmology.

%This is the easiest to explain in this picture, since the gravitational $\langle h  h\rangle$ correlators observed in the sky are dual to 
%energy-momentum tensor $\langle \delta T \delta T \rangle$ correlators on the boundary. So energy-momentum tensor (fluctuations)
%correlators are generated on the boundary, and by the AdS/CFT dictionary, the classical perturbations we see in bulk gravity now correspond to 
%quantum perturbations in the UV of the theory, so it is even more natural than in inflation now.

%6. {\bf Baryon asymmetry problem}
%
%This has the same explanation as in inflation, namely the same resolution as the entropy problem, being related to the same $S_1\sim 10^9$. 

\paragraph{\bf Conclusions} 
In this paper we have shown that the (non-geometric) holographic cosmology model of \cite{McFadden:2009fg} is capable of solving the standard problems of Hot Big Bang cosmology: the smoothness and horizon 
problems, the flatness problem and the monopole and relic problems.
In holographic cosmology time evolution translates into inverse RG flow and these problems are naturally resolved using properties of the RG flow. In these models the resolution of the initial singularity is mapped to the IR finiteness of the dual QFT and the arrow of time is linked with the monotonicity of RG flow. 
%Just like in the inflationary case, the source of the solution for most of the problems
%was the exponential dilution of perturbations during the $N_e$ e-folds of inflation, in holographic cosmology the source of 
%the solution is the fact that the energy-momentum tensor $T_{ij}$, coupling to tensor and scalar metric perturbations 
%$h_{ij}$, is a (marginally) relevant operator, that dilutes in the UV, dual to the future of the cosmology. In the case of 
%the monopole problem (as opposed to a generic relic problem), we needed in addition to show that $\tilde j_\mu^a$, the current 
%coupling to the monopole perturbation, is also (marginally) relevant, implying a dilution of the monopole gauge field 
%$A_\mu$ perturbation in the future of the cosmology. Also like in the case of inflation, we can use the solution of the 
%problems to constrain the parameters of the model: if in the case of inflation, it amounted to constraints on the number
%of e-folds until a certain dilution is achieved, in holographic cosmology we have constraints on the amount of 
%RG flow for the same.
Together with the previously found fact that the CMBR fitting is as good as for standard $\Lambda$CDM with inflation our results mean that holographic cosmology is a viable alternative for a Standard Model of Cosmology. 
%This should not be 
%seen as surprising, given the fact that the two models are related, as we have argued, by varying the strength of gravity, 
%or of the dual field theory.

%==============================================================================
\begin{acknowledgments}
%==============================================================================
\paragraph{Acknowledgments} We would like to thank Juan Maldacena and Paul McFadden for useful discussions. KS would like to the ICTP-SAIFT and USP for hospitality during the initial stage of this work, and acknowledges support from FAPESP grants 
nr. 2016/01343-7, and  nr. 2014/18634-9. This project has received funding/support from the European Union's Horizon 2020 research and innovation programme under the Marie Sklodowska-Curie grant agreement No 690575. KS is also supported in part by the Science and Technology Facilities Council (Consolidated Grant ``Exploring the Limits of the Standard Model and Beyond''). 
The work of HN is supported in part by CNPq grant 304006/2016-5 and FAPESP grant 2014/18634-9. HN would also like to 
thank the ICTP-SAIFR for their support through FAPESP grant 2016/01343-7. 

\end{acknowledgments}

\end{document}